\documentclass[aps,prd,twocolumn,superscriptaddress,showpacs,nofootinbib,altaffilletter]{revtex4-1}
\usepackage{amsmath,amssymb}\begin{document}
\title{$w$-cosmological singularities}
\author{L. Fern\'andez-Jambrina}
\email[]{leonardo.fernandez@upm.es}
\homepage[]{http://dcain.etsin.upm.es/ilfj.htm}
\affiliation{Matem\'atica Aplicada, E.T.S.I. Navales, Universidad
Polit\'ecnica de Madrid,\\
Arco de la Victoria s/n, \\ E-28040 Madrid, Spain}%
\date{\today}
\begin{abstract}
In this paper we characterize barotropic index singularities of 
homogeneous isotropic cosmological models \cite{wsing}. They are shown to appear 
in cosmologies for which the scale factor is analytical with a Taylor 
series in which the linear and quadratic terms are absent. Though the 
barotropic index of the perfect fluid is singular, the singularities 
are weak, as it happens for other models for which the density and 
the pressure are regular.
\end{abstract}

\maketitle

\section{Introduction}

The observational evidence from different sources \cite{snpion,
Davis:2007na,WoodVasey:2007jb,Leibundgut:2004,wmap} for the present
stage of accelerated expansion of our universe has driven the quest
for theoretical explanations of such feature.  Assuming the validity
of the theory of gravity, one attempt of explanation is the existence
of an unregarded, but dominant at present time, ingredient of the
energy content of the universe, known as dark energy
\cite{Padmanabhan:2006ag,Albrecht:2006um,Sahni:2006pa}, with unusual
physical properties.  The other possibility is modifying the general
theory of relativity at large scales
\cite{Maartens:2007,Durrer:2007re,Padmanabhan:2007xy}.

Both approaches have contributed to change our view of the final 
state of the universe. Before the discovery of the accelerated 
expansion of the universe only two possibilities were considered. 
Either our universe would expand forever or the matter content would 
force a contraction and recollapse of the universe in a final Big 
Crunch. 

Observations compatible with a barotropic index $w=p/\rho$ lower than 
-1 pointed out a final singularity in the form of an infinite scale 
factor of the universe, named as Big Rip \cite{Caldwell:2003vq}. 
Other models were postulated and the family of candidates increased. 
The price to pay was violation of one or several energy conditions 
and hence these possibilities were not considered in classical 
theorems of singularities \cite{HE}. Among these we may find:
\begin{itemize}
    \item Sudden singularities: Finite-time singularities for which
    the weak and strong energy conditions hold, but the pressure of
    the cosmological fluid blows up whereas the density remains finite
    \cite{sudden}.  If the second derivative of the scale factor is
    positive, they are called Big Boost singularities \cite{boost}.
    Related to braneworld models for which the embedding of the brane
    in the bulk is singular at some point they have also been named
    quiescent singularities \cite{quiescent}. However the name
    quiescent appeared originally in a different context in
    \cite{Andersson:2000cv} related to non-oscillatory singularities.

    \item Generalized sudden singularities: These are finite-time
    singularities with finite density and pressure \cite{tsagas} 
    instead of diverging pressure. Again in the braneworld context they 
    have also been called quiescent \cite{quiescent1}, though this 
    name had already been assigned to sudden singularities.

    \item  Big Brake: These singularities arise originally in 
    tachyonic models and are characterized by a negative infinite 
    second derivative of the scale factor whereas the first 
    derivative vanishes and the scale factor remains finite \cite{brake}. 
    They are 
    consequently a subcase of sudden singularities.

    \item  Big Freeze: These singularities were detected in 
    generalised Chaplygin models and are characterised by a finite 
    scale factor and an infinite density \cite{freeze}.

    \item Inaccesible singularities: These singularities appear in
    cosmological models with toral spatial sections, due to infinite 
    winding of trajectories around the tori. For instance, 
    compactifying spatial sections of the de Sitter model to cubic tori. 
    However, these singularities cannot be reached by physically 
    well-defined observers. This fact suggests the name of 
    inaccesible singularities  \cite{mcinnes}.
    
    \item Directional singularities: Curvature scalars vanish at the 
    singularity but there are causal geodesics along which the 
    curvature components diverge \cite{hidden}. That is, the singularity is 
    encountered just for some observers. In a general framework they 
    were dubbed p.p curvature singularities (curvature singularities 
    with respect to a parallelly propagated basis) in \cite{HE}.
\end{itemize}
    
Most of them are compiled in a classification due to 
Nojiri, Odintsov and Tsujikawa (N.O.T. in the following) in 
terms of which physical quantities blow up \cite{Nojiri:2005sx}:

\begin{itemize}    
   \item Big Bang / Crunch: Zero $a$, divergent $H$, density and 
   pressure.
   
   \item Type I: ``Big Rip'': Divergent $a$, density and pressure.

   \item Type II: ``Sudden'': Finite $a$, $H$, density, divergent 
   $\dot H$ and pressure. They enclose Big Brake and 
   most of quiescent singularities. 

   \item Type III: ``Big Freeze'' or ``Finite Scale Factor 
   singularities'': finite $a$, divergent $H$, 
   density and pressure.

   \item Type IV: ``Generalised sudden'': Finite $a$, $H$, $\dot H$, density, 
   pressure, divergent higher derivatives. They comprise the subcase 
   of quiescent singularities with finite pressure.
\end{itemize}

This classification is refined further in \cite{IV} and \cite{yurov}.

Since innaccesible and directional singularities are not in principle 
related to divergences in curvature scalars they would fall out of 
this scheme.

Some of these cannot be taken as the end of the universe, 
since the spacetime can be extended continuosly beyond the 
singularity \cite{suddenferlaz,puiseux,modigravi}. The case of a 
string surviving a sudden singularity is proven in \cite{string}.

In \cite{wsing} a cosmological model with just a singular barotropic
index at $t=t_{s}$ is described,
\begin{eqnarray}
\label{wmodel}
a(t)&=&\frac{a_s}{1-\frac{3\gamma}{2}\left(\frac{n-1}{n-\frac{2}{3\gamma}}\right)^{n-1}}\nonumber
\\&+&\frac{1-\frac{2}{3\gamma}}{n-\frac{2}{3\gamma}}\frac{na_s}{1-\frac{2}{3\gamma}\left(\frac{n-\frac{2}{3\gamma}}{n-1}\right)^{n-1}}\left(\frac{t}{t_s}\right)^{\frac{2}{3\gamma}} \nonumber
\\&+& \frac{a_s}{\frac{3\gamma}{2}\left(\frac{n-1}{n-\frac{2}{3\gamma}}\right)^{n-1}-1}\left(1-\frac{1-\frac{2}{3\gamma}}{n-\frac{2}{3\gamma}}\frac{t}{t_s}\right)^n~,
\end{eqnarray}
where $\gamma>0$, in order to prevent the model from becoming 
phantom, and $n\neq1$. The constant $\gamma=w-1$ is related to the 
barotropic index $w$ near the Big Bang at $t=0$. We shall use the 
subindex $s$ throughout the paper to refer to quantities calculated 
at the time of the singularity $t_{s}$.

The scale factor, $a(t_{s})=a_{s}$ is regular and the density and the
pressure vanish at $t_{s}$. Furthermore, if $n$ is natural, the 
derivatives of the Hubble parameter are regular either. 
However, the effective barotropic index $w$ is infinite at $t_{s}$.

In this paper we would like to characterize these $w$-singularities 
in FLRW cosmological models. 

In the next section we obtain the cases 
for which the barotropic index is singular and check which of them 
have vanishing fluid density and pressure at the singularity. 
Finally, the cases with singularities in higher derivatives of the 
scale factor are removed. A final section of conclusions is included.

\section{Singularities in barotropic index}
The total content of a FLRW spacetime is described as a perfect fluid 
of density $\rho$ and pressure $p$. Since both of them are functions 
of just the time coordinate, the fluid has at least locally an 
equation of state $p=p(\rho)$. The quotient of both is the barotropic 
index, $w=p/\rho$, which is also a function of time. Focusing 
on flat cosmologies, 
\begin{equation}
ds^2=-dt^2+a^2(t)\left\{dr^2+ r^2\left(d\theta^2+\sin^2\theta
d\phi^2\right)\right\},\label{metric}\end{equation} the barotropic
index is constant just for power-law flat, $w\neq-1$, and de Sitter
models, $w=-1$.

From Friedmann equations for the effective pressure and energy 
density 
\begin{equation}\label{flrw} H=\frac{\dot a}{a},\qquad
3H^2=\rho,\qquad \dot \rho+3H(\rho+p)=0,\end{equation}
we get the expression for the barotropic index $w$,
\begin{equation}
w=-\frac{1}{3}-\frac{2}{3}\frac{a\ddot a}{\dot a^2},\end{equation}
in terms of the derivatives of the scale factor $a(t)$.

Assuming that the scale factor admits a generalized power expansion 
\cite{visser,puiseux} 
of the form
\begin{eqnarray}\label{expan}
&&a(t)=c_{0}(t_{ s}-t)^{\eta_{0}}+c_{1}(t_{ s}-t)^{\eta_{1}}+\cdots,
\nonumber\\ &&\eta_{0}<\eta_{1}<\cdots,\qquad c_{0}>0,
\end{eqnarray}
around a value $t_{ s}$, with real exponents, we may expand the barotropic index 
accordingly:

\begin{itemize}
    
\item If $\eta_{0}\neq 0$,
\begin{eqnarray*}w&=&-\frac{1}{3}-\frac{2}{3}\left(\frac{\eta_{0}-1}{\eta_{0}}+
\frac{c_{1}\eta_{1}(\eta_{1}-1)}{c_{0}\eta_{0}^2}(t_{ s}-t)^{\eta_{1}-\eta_{0}}
+\cdots\right)\\&&
\left(1+\frac{c_{1}}{c_{0}}(t_{ s}-t)^{\eta_{1}-\eta_{0}}+\cdots\right)\\&&
\left(1-\frac{c_{1}\eta_{1}}{c_{0}\eta_{0}}(t_{ s}-t)^{\eta_{1}-\eta_{0}}+
\cdots\right)^2
\simeq-\frac{1}{3}-\frac{2}{3\eta_{0}}(\eta_{0}-1)\\&-&\frac{2c_{1}}{3c_{0}\eta_{0}}
\left(\eta_{0}-1+\frac{\eta_{1}(\eta_{1}-2\eta_{0}+1)}{\eta_{0}}\right)
(t_{ s}-t)^{\eta_{1}-\eta_{0}},\end{eqnarray*}
the result is obviously consistent at $t=t_{ s}$ with a linear 
barotropic perfect fluid, for which $\eta_{0}=2/3(1+w_{ s})$ with 
finite $w_{ s}$. In the limit of large $\eta_{0}$ de 
Sitter-like models would appear.

\item If $\eta_{0}= 0$, the expansion becomes more involved,
\begin{eqnarray*}w&=&-\frac{1}{3}-\frac{2}{3\eta_{1}}\left(\eta_{1}-1+
\frac{c_{2}\eta_{2}(\eta_{2}-1)}{c_{1}\eta_{1}}(t_{ s}-t)^{\eta_{2}-\eta_{1}}
+\cdots\right)\\&&
\left(\frac{c_{0}}{c_{1}}(t_{ s}-t)^{-\eta_{1}}+1+\frac{c_{2}}{c_{1}}(t_{ s}-t)^{\eta_{2}-\eta_{1}}+\cdots\right)\\&&
\left(1-\frac{c_{2}\eta_{2}}{c_{1}\eta_{1}}(t_{ s}-t)^{\eta_{2}-\eta_{1}}+
\cdots\right)^2
\simeq-\frac{1}{3}\\&-&\frac{2c_{0}}{3c_{1}\eta_{1}}(\eta_{1}-1)(t_{ s}-t)^{-\eta_{1}}\\&-&
\frac{2c_{0}c_{2}\eta_{2}}{3c_{1}^2\eta_{1}^2}
\left(\eta_{2}-2\eta_{1}+1\right)
(t_{ s}-t)^{\eta_{2}-2\eta_{1}}-\frac{2(\eta_{1}-1)}{3\eta_{1}},\end{eqnarray*}
since several possibilities arise: 

\begin{itemize}
\item If $\eta_{1}\neq 1$, the barotropic index diverges as a power 
$(t_{ s}-t)^{-\eta_{1}}$.

\item If $\eta_{1}=1$, depending on the value of $\eta_{2}$,
\[w\simeq -\frac{1}{3}-\frac{2c_{0}c_{2}\eta_{2}\left(\eta_{2}-1\right)}{3c_{1}^2}
(t_{ s}-t)^{\eta_{2}-2},\]
we may have a singular barotropic index for $\eta_{2}\in (1,2)$
and a regular one for $\eta_{2}>2$. The subcase $\eta_{2}=2$,
\begin{eqnarray*}w&\simeq &
-\frac{1}{3}-\frac{4c_{0}c_{2}}{3c_{1}^2}+\left(\frac{16c_{0}c_{2}^2}{3c_{1}^3}
-\frac{4c_{2}}{3c_{1}}\right)(t_{ s}-t)\\&-&\frac{2c_{0}c_{3}\eta_{3}}{3c_{1}^2}(t_{ s}-
t)^{\eta_{3}-2},\end{eqnarray*}
produces also a regular $w$ around $t_{ s}$.

\end{itemize}\end{itemize}

Models with scale factors admitting no generalized power series, 
typically models with $a(t)\sim e^{b/(t_{ s}-t)^p}$, $p>0$, produce finite 
barotropic indices of the form
\[w\sim -1-\frac{2}{3}\frac{p+1}{bp}(t_ s-t)^p,\]
and are therefore no candidates for producing $w$-singularities.

A directional singularity of the type of \cite{hidden} cannot be a 
$w$-singularity since the former has a finite barotropic index.

Therefore, the only chances for a diverging barotropic index arise 
for $\eta_{0}=0$, $\eta_{1}\neq1$ or $\eta_{0}=0$, $\eta_{1}=1$, 
$\eta_{2}<2$, as consigned in Table \ref{tablw}.

\begin{table}[h]
   \begin{tabular}{cccc}
   \hline
   ${\eta_{0}}$ & ${\eta_{1}}$ & $\eta_{2}$ &$w_{ s}$ \\
   \hline
   $\neq0$ & $(\eta_{0}, \infty)$ &   $(\eta_{1}, \infty)$ & Finite \\ 
   $0$ & $(0,1)$ &   $(\eta_{1}, \infty)$ &   Infinite  \\
      & $1$ &  $(1,2)$ & Infinite  \\
      & $1$ & $[2, \infty)$ & Finite  \\
      & $(1, \infty)$ & $[\eta_{1}, \infty)$ & Infinite \\
   \hline
   \end{tabular}
\caption{Singularities in  barotropic index}\label{tablw}
\end{table}

In order to get a $w$-singularity, besides a diverging barotropic 
index, we need vanishing density and pressure,
\begin{equation}
\rho=3\left(\frac{\dot a}{a}\right)^2,\qquad
p=-\left(\frac{\dot a}{a}\right)^2-\frac{2\ddot
a}{a}.\end{equation} 
We check these conditions for both singular cases:

\begin{enumerate}
    \item $\eta_{0}=0$, $\eta_{1}\neq1$: 
    $a(t)=c_{0}+c_{1}(t_{ s}-t)^{\eta_{1}}+\cdots$.
\begin{eqnarray*}
\rho&=&
\frac{3c_{1}^2\eta_{1}^2}{c_{0}^2}(t_{ s}-t)^{2(\eta_{1}-1)}+\cdots,\\
p&=&-
\frac{2c_{1}\eta_{1}(\eta_{1}-1)}{c_{0}}(t_{ s}-t)^{\eta_{1}-2}+\cdots.
\end{eqnarray*}

The expansions show that density tends to zero for 
$\eta_{1}>1$, whereas a vanishing pressure requires $\eta_{1}>2$.

We have then both vanishing density and pressure and divergent
barotropic index for $\eta_{0}=0$, $\eta_{1}>2$.

    \item $\eta_{0}=0$, $\eta_{1}=1$, $\eta_{2}<2$: 
    $a(t)=c_{0}+c_{1}(t_{ s}-t)+c_{2}(t_{ s}-t)^{\eta_{2}}+\cdots$.
\begin{eqnarray*}
\rho&=&
\frac{3c_{1}^2}{c_{0}^2}+
\frac{6\eta_{2}c_{1}c_{2}}{c_{0}^2}(t_{ s}-t)^{\eta_{2}-1}\\&-&
6\left(\frac{c_{1}}{c_{0}}\right)^3(t_{ s}-t)+\cdots
,\\
p&=&-
\frac{2c_{2}\eta_{2}(\eta_{2}-1)}{c_{0}}(t_{ s}-t)^{\eta_{2}-2}-
\frac{c_{1}^2}{c_{0}^2}+\cdots.
\end{eqnarray*}

Since the density is finite and the pressure diverges in this case, it cannot be a 
$w$-singularity, but a sudden singularity.
\end{enumerate}

For vanishing pressure and density and divergent barotropic index we 
are left just with the $\eta_{0}=0$, $\eta_{1}>2$ case:

\textit{A FLRW cosmological model has a singular barotropic 
index $w$ with vanishing pressure and density at a finite
time $t_{ s}$ if and only if the generalized power expansion of
the scale factor $a(t)$ is of the form}
\begin{equation} \label{wnon}
    a(t)=c_{0}+c_{1}(t_{ s}-t)^{\eta_{1}}+\cdots,\end{equation}
\textit{with $\eta_{1}>2$.}

\textit{If we allow finite pressure, the 
condition is relaxed to $\eta_{1}>1$.}

Finally, since the scale factor does not vanish at $t_{s}$ the only 
possibility for a singularity in higher derivatives of the Hubble 
factor is that a derivative of the scale factor (7) blows up. If 
$\eta_{1}$ is non-integer, there will be derivatives
$a^{p)}(t)\sim c_{1}(t_{s}-t)^{\eta_{1}-p}$ which blow up for 
$p>\eta_{1}$. 

The only way to prevent this is to require that 
$\eta_{1}$ be natural. But then the reasoning would be the same for 
$\eta_{2}$ and the subsequent exponents. Hence, the only possibility to avoid a diverging 
derivative of the scale factor is that \emph{every} exponent 
$\eta_{i}$ be natural. But in this case the series is no longer a 
generalised power series, but a Taylor series. Since $\eta_{1}>2$, 
the lowest power would be at least three:

\textit{A FLRW cosmological model has a $w$-singularity
at a finite time $t_{ s}$ if and only if the scale factor $a(t)$ 
admits a Taylor series at $t_{s}$ with vanishing linear and quadratic 
terms,}
\begin{equation}a(t)=c_{0}+\sum_{n=3}^{\infty}c_{n}(t_{ 
s}-t)^{n}.\end{equation}
\textit{If we allow finite pressure, then just 
the linear term is to vanish.}
\section{Discussion \label{discuss}}

Cosmological models with generalized power expansions of the scale 
factor have been discussed in \cite{puiseux}. The exponents of the 
power expansion are related to the appearance of cosmological 
singularities, which can be strong or weak. 

Weak singularities are not actual singularities in the sense that the
spacetime can be extended continuously beyond the singularity.  Or,
put in another way, from the physical point of view, a finite object
is not necessarily crushed on crossing a weak singularity.  The
classification of singularities \cite{puiseux} in terms of the
exponents of the scale factor expansion is recorded in Table
\ref{scale}. 

The column $\{\eta_{i}\}$ stands for the properties of 
the exponents of the expansion: I means no additional condition on 
them, S means that at least one exponent must be non-natural in order to 
have a singularity in one of the derivatives and N means that every 
exponent is natural.
\begin{table}
     \begin{tabular}{ccccccc}
   \hline
   ${\eta_{0}}$ & ${\eta_{1}}$ & $\eta_{2}$ & $\{\eta_{i}\}$ &\textbf{Tipler} &
   \textbf{Kr\'olak} & \textbf{N.O.T.} \\
   \hline
   $(- \infty,0)$ & $(\eta_{0}, \infty)$ &   $(\eta_{1}, \infty)$ & I&
   Strong & Strong  & I\\ 
   $0$ & $(0,1)$ &   $(\eta_{1}, \infty)$ & S&  Weak & Strong & III \\
      & $1$ & $(1,2)$ & S& Weak & Weak & II \\
      &  & $[2, \infty)$ & S & Weak & Weak & IV \\
      & $(1,2)$ &  $(\eta_{1}, \infty)$ & S & Weak & Weak & II \\
      & $[2, \infty)$ &   $(\eta_{1}, \infty)$ & S&  Weak & Weak
      &  IV \\      & $[3, \infty)$\footnote{If we include finite 
      pressure $w$-singularities, 
       then $\eta_1\in[2,\infty)$.} &   $[\eta_{1}+1, \infty)$ & N & Weak & Weak
      &  $w$ \\ $(0, \infty)$ & $(\eta_{0}, \infty)$ &   $(\eta_{1}, 
      \infty)$ & I&
Strong & Strong & Crunch\\
   \hline
   \end{tabular}
\caption{Singularities in  cosmological models}\label{scale}
\end{table}

The difference between Tipler's \cite{tipler} and
Kr\'olak's \cite{krolak} criterion for the strength of singularities
is just that, whereas the former requires the volume of finites objects
to tend to zero at a strong singularity, the latter just imposes the
derivative of the volume with respect to proper time to be negative, 
which is a milder requirement. 
Conditions for checking both criteria may be found in \cite{clarke}. 
Another criterion is the one in \cite{rudnicki}.

All cosmological models with $w$-singularities therefore belong to the 
last but one line of the classification and hence we may conclude 
that $w$-singularities are weak singularities. 


Therefore, the diverging barotropic index for $w$-singularities, 
which is not shared necessarily by Type IV singularities, does not 
influence the weak character of both families of singularities.

\section*{Acknowledgments}L. F.-J. wishes to
thank the University of the Basque
Country for their hospitality and facilities to carry out this work. 
The author wishes to thank the referees for their useful comments.


\begin{thebibliography}{99}
\bibitem{wsing} M.P. D\c abrowski, T. Denkiewicz, \emph{Phys. Rev. D} 
\textbf{79}, 063521 (2009).
\bibitem{snpion}
A. G. Riess et al. [Supernova Search Team Collaboration],
Astron. J. {\bf 116} (1998)  1009  [arXiv:astro-ph/9805201];
S. Perlmutter et al. [Supernova Cosmology Project Collaboration],
Astrophys. J. 517, 565 (1999)
[arXiv:astro-ph/9812133].
\bibitem{Davis:2007na}
 T.~M.~Davis {\it et al.},
 Astrophys.\ J.\  {\bf 666} (2007) 716
 [arXiv:astro-ph/0701510].
\bibitem{WoodVasey:2007jb}
 W.~M.~Wood-Vasey {\it et al.}  [ESSENCE Collaboration],
 Astrophys.\ J.\  {\bf 666} (2007) 694
 [arXiv:astro-ph/0701041].
\bibitem{Leibundgut:2004} B. \,Leibundgut, 
in Reviews of Modern Astronomy  {\bf 17} (2004)
edited by R. E. Schielicke (Wiley-VCH, Weinheim)
\bibitem{wmap}  D.~N.~Spergel {\it et al.}  [WMAP Collaboration],
 Astrophys.\ J.\ Suppl.\  {\bf 148} (2003) 175
 [arXiv:astro-ph/0302209];
 D.~N.~Spergel {\it et al.}  [WMAP Collaboration],
 Astrophys.\ J.\ Suppl.\  {\bf 170} (2007) 377
 [arXiv:astro-ph/0603449];
 J.~Dunkley {\it et al.}  [WMAP Collaboration],
Observations:
 arXiv:0803.0586 [astro-ph], E.~Komatsu {\it et al.}  [WMAP Collaboration],
arXiv:0803.0547 [astro-ph].
\bibitem{Padmanabhan:2006ag}
T.~Padmanabhan,
 AIP Conf.\ Proc.\  {\bf 861} (2006) 179
 [arXiv:astro-ph/0603114].
\bibitem{Albrecht:2006um}
 A.~Albrecht {\it et al.},
 arXiv:astro-ph/0609591.
\bibitem{Sahni:2006pa}
 V.~Sahni and A.~Starobinsky,
 Int.\ J.\ Mod.\ Phys.\  D {\bf 15} (2006) 2105
 [arXiv:astro-ph/0610026].
\bibitem{Maartens:2007} Roy Maartens, J. Phys.: Conf. Ser. 68 (2007)
012046.
\bibitem{Durrer:2007re}
 R.~Durrer and R.~Maartens,
 Gen.\ Rel.\ Grav.\  {\bf 40} (2008) 301
 [arXiv:0711.0077 [astro-ph]].
\bibitem{Padmanabhan:2007xy}
 T.~Padmanabhan,
 arXiv:0705.2533 [gr-qc].
\bibitem{Caldwell:2003vq}
 R.~R.~Caldwell, M.~Kamionkowski and N.~N.~Weinberg,
 Phys.\ Rev.\ Lett.\  {\bf 91} (2003) 071301
 [arXiv:astro-ph/0302506].
 \bibitem{HE} S.W. Hawking, G.F.R. Ellis,
\textit{The Large Scale Structure of Space-time}, Cambridge University
Press, Cambridge, (1973).
\bibitem{sudden} J.D. Barrow, \textit{Class. Quant. Grav.} {\bf 21}, L79 (2004)
;  S. Nojiri, S.D. Odintsov, \textit{Phys.\ Lett. B} 
 {\bf 595}, 1 (2004); 
J.D. Barrow,
  \textit{Class.\ Quant.\ Grav.}\  {\bf 21}, 5619 (2004); 
K. Lake,
  \textit{Class.\ Quant.\ Grav.}  {\bf 21}, L129 (2004);  
 S. Nojiri, S.D. Odintsov, \textit{Phys.\ 
  Rev.\ D} {\bf 70}, 103522 (2004); 
  M.P. D\c abrowski,
  \textit{Phys.\ Rev.\ D} {\bf 71}, 103505 (2005); 
L.P. Chimento,  R. Lazkoz,
\textit{Mod.\ Phys.\ Lett.\ A} {\bf 19}, 2479 (2004) ; 
 M.P. D\c abrowski,
\textit{Phys.\ Lett.\ B} {\bf 625}, 184 (2005); 
J.D. Barrow, A.B. Batista, J.C. Fabris, S. Houndjo, \emph{Phys. Rev. 
D} \textbf{78}, 123508 (2008); J.D. Barrow, S.Z.W. Lip, \emph{Phys. 
Rev. D} \textbf{80}, 043518 (2009); S. Nojiri, S.D. Odintsov,
\emph{Phys. Rev. D} \textbf{78}, 046006 (2008); 
J.D. Barrow, S. Cotsakis, A. Tsokaros, \emph{Class. Quant. Grav.} 
\textbf{27}, 165017 (2010); 
J.D. Barrow, S. Cotsakis, A. Tsokaros, [arXiv:1003.1027] (2010).
\bibitem{boost} A.O. Barvinsky, C. Deffayet, A.Yu. Kamenshchik, 
\emph{JCAP} \textbf{05}, 034 (2010) [arxiv:0801.2063].
\bibitem{quiescent} Y. Shtanov, V. Sahni, \emph{Class. Quant. Grav.}
 \textbf{19}, L101 (2002) [arXiv:gr-qc/0204040].
\bibitem{Andersson:2000cv} L.~Andersson and A.~D.~Rendall,
 Commun.\ Math.\ Phys.\  {\bf 218} (2001) 479
 [arXiv:gr-qc/0001047].
\bibitem{tsagas} J.D. Barrow, C.G. Tsagas, 
\textit{Class.\ Quant.\ Grav.}\  {\bf 22}, 1563 (2005).
 
\bibitem{quiescent1} P. Tretyakov, A. Toporensky, Y. Shtanov, V. Sahni,
 \emph{Class. Quant. Grav.} \textbf{23}, 3259 (2006)
 [arXiv:gr-qc/0510104].


\bibitem{brake} V. Gorini, A.Y. Kamenshchik, U. Moschella, 
V. Pasquier,
 \emph{PRD} {\bf 69}, 123512 (2004).
\bibitem{freeze} M.~Bouhmadi-L\'opez, P.~F.~Gonzalez-D\'\i az and
P.~Mart\'\i n-Moruno,
 \emph{Phys.\ Lett.\  B} {\bf 659}, 1 (2008).
\bibitem{mcinnes} B. McInnes, \textit{Class. Quant. Grav.} 
\textbf{24}, 1605 (2007).
\bibitem{hidden}L.~Fern\'andez-Jambrina, \emph{
Phys.\ Lett.\ B} \textbf{656}, 9 (2007) [arXiv:gr-qc/0704.3936].
\bibitem{Nojiri:2005sx}
 S.~Nojiri, S.~D.~Odintsov and S.~Tsujikawa,
 Phys.\ Rev.\  D {\bf 71} (2005) 063004
 [arXiv:hep-th/0501025].
\bibitem{IV} M.P. D\c browski, T. Denkiewicz, [arXiv:0910.0023].
\bibitem{yurov} A.V. Yurov, \emph{Phys. Lett. B} \textbf{689}, 1 
(2010) [arxiv:0905.1393].
\bibitem{suddenferlaz}  L.~Fern\'andez-Jambrina and R.~Lazkoz,
 Phys.\ Rev.\  D {\bf 70}, 121503 (2004)
 [arXiv:gr-qc/0410124].
\bibitem{puiseux}
 L.~Fern\'andez-Jambrina and R.~Lazkoz,
 Phys.\ Rev.\  D {\bf 74}, 064030 (2006)
 [arXiv:gr-qc/0607073].
\bibitem{modigravi}L.~Fern\'andez-Jambrina, R. Lazkoz, \emph{Phys. Lett. 
B} \textbf{670}, 254-258 (2009) [arXiv:0805.2284].
\bibitem{string}  A. Balcerzak, M.P. D\c abrowski,
\textit{Phys.\ Rev.\ D} {\bf 73}, 101301 (2006).
\bibitem{visser} C. Catto\"en and  M. Visser,
Class.\ Quant.\ Grav.\  {\bf 22} (2005) 4913
[arXiv:gr-qc/0508045].
\bibitem{tipler} F.J. Tipler, {Phys. Lett.} \textbf{A64}, 8 (1977). 
\bibitem{krolak} A. Kr\'olak, Class. Quant. Grav. \textbf{3},
267 (1986). 
\bibitem{clarke} C.J.S. Clarke and A. Kr\'olak, Journ. Geom. Phys.
\textbf{2}, 127 (1985). 
\bibitem{rudnicki} W.~Rudnicki, R.~J.~Budzynski and W.~Kondracki,
Mod.\ Phys.\ Lett.\ A {\bf 21}, 1501 (2006).
[arXiv:gr-qc/0606007].
\end{thebibliography}
\end{document}